\documentclass[aps,preprintnumbers,twocolumn,showpacs]{revtex4-1}
\usepackage{latexsym}
\usepackage{amsfonts,amssymb,amsmath}
\usepackage{graphics,epsfig,placeins,subfigure,wrapfig}
\usepackage{hyperref}

 \newcommand{\bq}{\begin{equation}}
 \newcommand{\eq}{\end{equation}}
 \newcommand{\bqn}{\begin{eqnarray}}
 \newcommand{\eqn}{\end{eqnarray}}
 \newcommand{\nb}{\nonumber}
 \newcommand{\lb}{\label}
\begin{document}
\preprint{YITP-18-44, IPMU18-0083}
\title{Gravitational collapse and formation of universal horizons in Einstein-$\mbox{\ae}$ther theory}
\author{Madhurima Bhattacharjee$^{1, 2}$, Shinji Mukohyama$^{3, 4}$, Mew-Bing Wan$^{1}$, Anzhong Wang $^{1,2}$}
\affiliation{$^{1}$ Institute for Advanced Physics and Mathematics, Zhejiang University of Technology, Hangzhou, 310032, China}
\affiliation{$^{2}$ GCAP-CASPER, Physics Department, Baylor University, Waco, TX 76798-7316, USA}
\affiliation{$^{3}$ Center for Gravitational Physics, Yukawa Institute for Theoretical Physics, Kyoto University, 606-8502, Kyoto, Japan}
\affiliation{$^{4}$ Kavli Institute for the Physics and Mathematics of the Universe (WPI),
The University of Tokyo Institutes for Advanced Study,
The University of Tokyo, Kashiwa, Chiba 277-8583, Japan}

\date{\today}

\begin{abstract}
We numerically study the gravitational collapse of a massless scalar field with spherical symmetry in Einstein-$\mbox{\ae}$ther theory, 
and show that apparent, spin-0 and dynamical universal horizons (dUHs) can be all formed. The spacetime and the $\mbox{\ae}$ther 
field are well-behaved and regular, including regions nearby these horizons (but away from the center of spherical symmetry). The 
spacetime outside the apparent and spin-0 horizons settles down to a static configuration, and some of such resulting static black holes 
were already found numerically in the literature. On the other hand, the proper distance of the outermost dUH from the apparent (or spin-0) 
horizon keeps increasing on $\mbox{\ae}$ther-orthogonal time slices. This indicates that the outermost dUH is evolving into the causal 
boundary, even for excitations with large speeds of propagation.

\end{abstract}


\maketitle

\section{Introduction}

Lorentz invariance (LI) is one of the fundamental symmetries of modern physics and strongly supported by observations  \cite{ KR11}. In fact,  the experiments carried out so far are all consistent with it, and there is no observational evidence to show that such a symmetry must be  broken at certain energy scales, although  constraints of such violations in the gravitational sector  are much weaker than those in the matter sector \cite{Mattingly05}. 

There are still various reasons to construct gravitational theories with broken LI. In particular, if space and time are quantized at the Planck scale, as we understand from the point of view of quantum gravity \cite{Kiefer12},  then LI cannot be a fundamental continuous  symmetry, but instead  should be an emergent one at low energies. Another motivation comes from modification of gravity at long distances to explain the accelerated expansion of the late-time universe. Following these lines of argument, Lorentz violating (LV) theories of gravity have attracted lots of interest in recent years. These include  ghost condensation \cite{ArkaniHamed:2003uy}, Einstein-$\mbox{\ae}$ther theory ($\mbox{\ae}$-theory) \cite{Jacobson} and   Ho\v{r}ava   gravity \cite{Horava}.

However, once LI is broken, different species of particles can travel with different (sometimes arbitrarily large) speeds. This suggests that black holes may exist only at low energies.  At high energies, signals with sufficiently large speeds initially emanated inside an event horizon (EH) can escape to infinity. However, in contrast to this physical  intuition,  it was found that there still exist absolute causal boundaries, the so-called {\em universal horizons} (UHs), and particles even with infinitely large velocities would just move along these boundaries and cannot escape to infinity    \cite{BS11,BJS11,BBM12}. This is closely related to the causality in LV theories of gravity. Since now the speeds of particles can be arbitrarily large, similar to Newton's theory, to preserve the causality, it is necessary to introduce a scalar field with globally timelike gradient, the so-called {\em khronon}, which  defines an absolute time, and all particles are assumed to move along its increasing direction, so the causality in the sense of the past and the future is assured (Cf. Fig.~1 in  \cite{GLLSW}).  Then, in  asymptotically flat  stationary spacetimes, there might exist a surface on which the timelike translation Killing vector  becomes orthogonal to the gradient of the khronon  (See e.g. Fig.~2 in  \cite{LSW16}). Hence, a particle must cross this surface and move inevitably  inward (towards the increasing direction of the  khronon), once it arrives at it, no matter how large its speed is. This is a one-way membrane, and particles even with infinitely large speed cannot escape from it, once they are trapped inside. So, it acts as an absolute horizon to all particles. UHs have been extensively studied  {  (see e.g. \cite{Wang17} and references therein),} including their thermodynamics \cite{BBM13,CLMV,DWWZ}. 
 
 In general relativity (GR), it is well known that EHs can be formed from gravitational collapse of realistic matter, which implies that black holes with EHs as their boundaries exist in our Universe. However, in LV theories since particles with speeds larger than that of light exist, such particles can cross them and escape  to infinity,  even initially they are trapped inside  EHs. So, EHs in such theories are no longer the one-way membranes.  Instead, now the black hole boundaries are replaced by UHs, as  argued above. Therefore, from the same astrophysical considerations as in GR, a key issue  is whether UHs can be also formed from gravitational collapse in our Universe    \cite{SAM14,TWSW15,BCCS16}. In this paper, we shall address this important  issue in the framework of $\mbox{\ae}$-theory, which propagates three kinds of modes, the usual spin-2 graviton plus the spin-1 and spin-0 ones \cite{Jacobson}. We numerically show the formation of dynamical UHs (dUHs), the generalization of UHs to dynamical spacetimes with spherical symmetry. We also find that the proper distance of the outermost dUH from the apparent (or spin-0) horizon keeps increasing on $\mbox{\ae}$ther-orthogonal time slices. To our best knowledge, this is the first time  to show explicitly that dUHs can be formed from gravitational collapse.

\section{${\ae}$-Theory and Spherical Collapse}

The fundamental variables of the gravitational sector of $\mbox{\ae}$-theory are $(g_{\mu\nu}, u^{\mu}, \lambda)$, where $g_{\mu\nu}$ is  the spacetime metric with the signatures $(-, +,+,+)$,  and $u^{\mu}$ is the aether four-velocity, while $\lambda$ is a Lagrangian multiplier, which guarantees that $u^{\mu}$ is always timelike and has unit norm. The general action of $\mbox{\ae}$-theory  takes the form  \cite{Jacobson}, $S = S_{\mbox{\ae}} + S_{m}$, where  $S_{\mbox{\ae}} \; (S_{m})$ denotes the action of gravity (matter),  given  by
\begin{eqnarray}
\label{2.1}
 S_{\mbox{\ae}} &=& \frac{1}{16\pi G_{\mbox{\ae}}}\int{\sqrt{- g} \; d^4x \Big[R + {\cal{L}}_{\mbox{\ae}}\left(g_{\mu\nu}, u^{\nu}, \lambda\right)\Big]},\nonumber\\
 S_{m} &=& \int{\sqrt{- g} \; d^4x \Big[{\cal{L}}_{m}\left(g_{\mu\nu}, \psi\right)\Big]}.
\end{eqnarray}
Here $G_{\mbox{\ae}}$ is related to  the Newtonian constant $G_{N}$ \cite{CL04} by 
$G_{N} = {G_{\mbox{\ae}}}/{(1 - c_{14}/2)}$,with  $c_{ij} \equiv c_i + c_j$ and $c_{ijk} = c_i + c_j + c_k$, and $\psi$ collectively denotes the matter fields. $R$  is  the  Ricci scalar,
 and ${\cal{L}}_{\mbox{\ae}}  \equiv - M^{\alpha\beta}_{~~~~\mu\nu}\left(D_{\alpha}u^{\mu}\right) \left(D_{\beta}u^{\nu}\right) + \lambda \left(g_{\alpha\beta} u^{\alpha}u^{\beta} + 1\right)$,
 where $D_{\mu}$ denotes the covariant derivative of $g_{\mu\nu}$, 
$M^{\alpha\beta}_{~~~~\mu\nu} \equiv c_1 g^{\alpha\beta} g_{\mu\nu} + c_2 \delta^{\alpha}_{\mu}\delta^{\beta}_{\nu} +  c_3 \delta^{\alpha}_{\nu}\delta^{\beta}_{\mu} - c_4 u^{\alpha}u^{\beta} g_{\mu\nu}$,
and $c_i$'s are four  independent dimensionless coupling constants. 

Recently,  the combination of the gravitational wave GW170817 \cite{GW170817}
and the  gamma-ray burst  GRB 170817A \cite{GRB170817} events provided  a remarkably stringent constraint on the speed of the spin-2 graviton, $- 3\times 10^{-15} < c_T -1 < 7\times 10^{-16}$. In  $\mbox{\ae}$-theory,  this implies $\left |c_{13}\right| < 10^{-15}$ \cite{JM04}. Together with other observational and theoretical constraints,  the parameter space of $\mbox{\ae}$-theory is restricted to the intersection of \cite{OMW18},
 \begin{eqnarray}
\label{2.2}
&&  \left|c_{13}\right| < 10^{-15}, \quad 0 \leq c_{14} \leq 2.5\times 10^{-5}, \nonumber\\
&&  0 \leq c_2 \leq 0.095,\quad c_{4} \leq 0\,.
 \end{eqnarray}
The variations of the total action with respect to $g_{\mu\nu}$ and  $u^{\mu}$ yield
 \begin{eqnarray}
 \label{2.3a}
&&  R^{\mu\nu} - \frac{1}{2}g_{\mu\nu}R - T^{\mu\nu}_{\mbox{\ae}} = 8\pi G_{\mbox{\ae}} T_{m}^{\mu\nu},\\
 \label{2.3b}
&&   D_{\alpha} J^{\alpha}_{~~~\mu} + c_4 a_{\alpha} D_{\mu}u^{\alpha} + \lambda u_{\mu} = 0,
 \end{eqnarray}
while its variation with respect to $\lambda$ yields $u^{\alpha}u_{\alpha} = -1$. Here, $T_{m}^{\mu\nu}$ denotes the matter energy-stress tensor,  and 
  $T_{\mbox{\ae}}^{\alpha\beta} \equiv
  -D_{\mu}\big[u^{(\beta}J^{\alpha) \mu} - J^{\mu(\alpha}u^{\beta)} - J^{(\alpha\beta)}u^{\mu}\big]
 - c_1\big[\left(D_{\mu}u^{\alpha}\right)\left(D^{\mu}u^{\beta}\right) - \left(D^{\alpha}u_{\mu}\right)\left(D^{\beta}u^{\mu}\right)\big]
 + c_4 a^{\alpha}a^{\beta}    + \lambda  u^{\alpha}u^{\beta} - \frac{1}{2}  g^{\alpha\beta} J^{\delta}_{\;\;\sigma} D_{\delta}u^{\sigma}$,
 with
$ J^{\alpha}_{\;\;\;\mu} \equiv M^{\alpha\beta}_{~~~~\mu\nu}D_{\beta}u^{\nu}\,$,
$a^{\mu} \equiv u^{\alpha}D_{\alpha}u^{\mu}$,
$\lambda = u_{\beta}D_{\alpha}J^{\alpha\beta} + c_4 a^2$, and $a^{2}\equiv a_{\lambda}a^{\lambda}$.

Gravitational collapse of a spherical massless scalar field in $\mbox{\ae}$-theory was already studied in some detail \cite{GEJ07,AGG18}. In particular, it was shown that  for two different sets of $c_i$'s [given, respectively, by Eqs.(16) and (34) with $c_1 = 0.7$ in  \cite{GEJ07}, which will be referred as to GEJ1 and GEJ2],   both  apparent horizons  (AHs) and spin-0 horizons (S0Hs) are formed during the collapse \cite{GEJ07}, and  the configurations finally settle down to  the regular static black holes found numerically  in  \cite{EJ06}. For another set of $c_i$'s, the collapse  instead  results in the temporary formation of a white hole horizon \cite{AGG18}, although the corresponding  static black hole exists   \cite{BJS11}. It should be noted that neither GEJ1 nor GEJ2 satisfies  the constraints (Eq.~\ref{2.2}). 

\begin{figure}
\includegraphics[width=\linewidth]{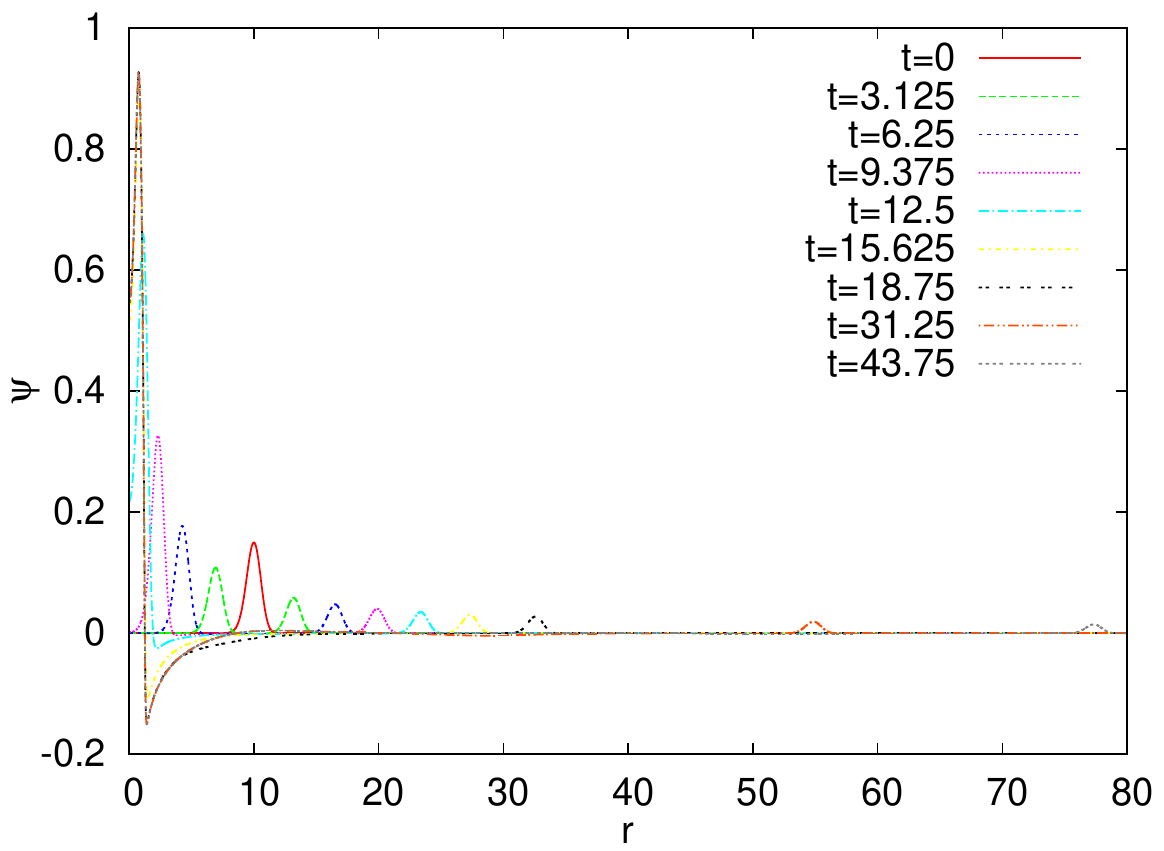}%
\caption{Evolution of the scalar field profile, $\Psi$ for the case GEJ1, using a medium-resolution simulation.   \label{fig:scalar}}
\end{figure}

Therefore, in this paper our goals are two-fold: First we show that even within the range of the new constraints, AHs and S0Hs can be still formed from gravitational collapse. Second, dynamical UHs can be also formed. To these goals, we  choose to study the same setup as that studied in  \cite{GEJ07,AGG18}, closely following their notation and conventions. This will in particular allow us to check our numerical codes. We choose the surfaces of constant time orthogonal to $u^{\mu}$ and the gauge that leads to  the form of metric, $ds^2 = \gamma_{ab}dx^adx^b + \Phi^2\left(d\theta^2 + \sin\theta^2 d\varphi^2\right)$, 
where $\gamma_{ab} dx^a dx^b = - \alpha^2 dt^2 + \left( dr + \beta^r dt \right)^2$, $a,b=0,1$; $\alpha,\; \beta^r$ and $\Phi$ are functions of $x^a=(t,r)$ only; and $u_{\mu}dx^{\mu} = u_adx^a = -\alpha dt$, for which the time evolution vector is given by $t^{\mu} = \alpha u^{\mu} + \beta^{\mu}$ with $\beta^{\mu} \partial_{\mu} = \beta^r \partial_r$. For the massless scalar field $\chi$ we have ${\cal{L}}_{m} =  - D_{\nu}\psi D^{\nu}\psi/(16\pi G_{\mbox{\ae}})$, where $\psi \equiv \sqrt{8\pi G_{\mbox{\ae}}}\; \chi$. The evolved quantities are then $(\psi, P, K, a_r, \Phi)$, where $P \equiv {\cal{L}}_u \psi$, and $K$ is the trace of the extrinsic curvature of constant-$t$ surfaces. The dynamical equations and  constraints are given, respectively, by 
%Eqs.(22)-(26) and Eqs.(27)-(30) in   
\cite{GEJ07},
\bqn
\lb{eq8}
\dot\psi &=& \alpha P + \beta^r \psi',\\
\lb{eq9}
\dot{P}&=& \beta^r P' + \alpha\left(PK + a^r \psi' + \psi'' + \frac{2\Phi'}{\Phi}\psi'\right),\\
\lb{eq10}
\dot{K} &=& \beta^r K' + \frac{\alpha}{3} K^2 + \frac{\alpha}{\Delta}\Bigg[2P^2 + 3\left(1-c_{13}\right)Q^2\nb\\
&&  +\left(c_{14} - 2\right)\left(a_r' + 2a_r \frac{\Phi'}{\Phi} + a_r^2\right)\Bigg],\\
\lb{eq11}
\dot{a}_r &=& \beta^ra_r' + \alpha\Bigg[\left(\frac{2K}{3} - Q\right)a_r + \frac{c_{13}}{c_{14}\left(1-c_{13}\right)}P\psi'\nb\\
&& ~~~~~~~~~~~~~~~ - \frac{c_{123}}{c_{14}\left(1-c_{13}\right)}K'\Bigg],\\
\lb{eq12}
\dot{\Phi} &=& \beta^r \Phi' + \alpha\Phi\left(\frac{Q}{2} - \frac{K}{3}\right),
\eqn
and 
\bqn
\lb{eq13}
Q' &=& -3Q\frac{\Phi'}{\Phi} + \frac{1}{1-c_{13}}\left( \frac{\Delta}{3} K' - P\psi'\right),  \\
\lb{eq14}
\frac{\alpha'}{\alpha} &=& a_r, \\
\lb{eq15}
{\beta^r}' &=& \alpha\left(Q + \frac{K}{3}\right),\\
\lb{eq16}
{\cal{C}} &=& \Phi'' + \frac{{\Phi'}^2 -1 }{2\Phi} + c_{14}a_r \Phi' + \frac{\Phi}{4}\Bigg[c_{14}\left(2a_r' + a_r^2\right)  \nb\\
&& + P^2 + {\psi'}^2 
+ \frac{3}{2}\left(1-c_{13}\right)Q^2 - \frac{\Delta}{3} K^2\Bigg] =  0,
\eqn
with   $Q \equiv K^r_r - K/3$, $\Delta \equiv 2 + c_{13} + 3c_2$, $\dot{\psi} \equiv \partial_t\psi$, $\psi' \equiv \partial_r\psi$, and so on.

The locations of the S0Hs and AHs are defined, respectively, by $\tilde{\gamma}^{ab} n_a n_b  = 0$ and $\gamma^{ab} n_a n_b = 0$, where $n_a \equiv \partial_a\Phi$, $\tilde{\gamma}^{ab} = (\tilde{\gamma}^{-1})^{ab}$ and $\tilde{\gamma}_{ab} \equiv \gamma_{ab} + (1-c_S^2) u_a u_b$ with $c_S^2 \equiv c_{123}(2-c_{14})/[c_{14}(1-c_{13})(2+c_{13} + 3c_2)]$ \cite{GEJ07}. Hereafter, by a S0H/AH we shall denote an outer S0H/AH. In stationary spacetimes, UHs are defined by $u_a\zeta^a = 0$, where $\zeta^a\partial_a$ is the time translation Killing vector \cite{BS11,Wang17}. However, when spacetimes are dynamical, such a  vector does not exist any longer. Following   \cite{TWSW15,Wang17} in defining a  dUH, we first introduce the Kodama vector \cite{Kodama80} (See also  Refs.~\cite{Hayward:1998ee,Abreu10}), {  $k^a \equiv \epsilon^{ab}_{\bot}n_b = \left(-\Phi_{,r}, \Phi_{,t}\right)/\alpha$, } where $ \epsilon^{ab}_{\bot}$  is the Levi-Civita tensor with $\epsilon^{01}_{\bot} = - 1 / \sqrt{-\gamma}$. It is clear that $k^a n_a  = 0$. For spacetimes that are asymptotically flat there always exists a region with sufficiently large $\Phi$, in which $n_a$ ($k^a$) is spacelike (time-like). An AH may form, say, at $r = r_{\text{AH}}$, where $n_a $ becomes null. Then, in the trapped region with $\gamma^{ab} n_a n_b < 0$,  $n_a \; (k^a)$ becomes timelike (spacelike).  We define the location of a  dUH as the surface at which 
\begin{equation}
\label{2.4}
u_a k^a  = 0,
\end{equation}
where in the current case  $u_a k^a =  \Phi_{, r}$.  Since $u_a$ is  globally timelike, Eq.(\ref{2.4}) is possible only when $k^a$ is spacelike. Clearly, this can be true only inside AH, that is, we must have $r_{\text{dUH}} < r_{\text{AH}}$. Eq.(\ref{2.4}) may have multiple roots, and what is relevant is the outermost dUH, i.e. the one with the largest $r$ (but not necessarily with the largest $\Phi$).  For the outermost dUH, we have ${\Phi}_{\text{dUH}} < {\Phi}_{\text{AH}}$ since $\Phi_{,r}>0$ for $r>r_{\text{dUH}}$. In the stationary spacetimes, the Kodama vector coincides with the time translation vector, and the above definition reduces to  static spacetimes, and later generalized to various stationary spacetimes {  (See \cite{Wang17} and references therein). }

\section{Numerical Setup and Results}

Our simulations are performed with a finite-differencing code. The initial data, numerical schemes and boundary conditions used in our code also closely follow   \cite{GEJ07}. The set of PDEs are solved on a uniformly spaced $r$-domain, where $r$ is the proper radial coordinate spanning $[0,r_{\text{max}}], \quad r_{\text{max}} = 80$ (or $=320$) with a spacing of $\Delta r=0.003125, 0.00625, 0.0125$, the high, medium and low resolutions, respectively. The timestep size is set to $0.2\times\Delta r$. In our code, the dynamical variables are integrated in time using an iterated Crank-Nicholson scheme with two iterations. We apply the 4th-order Kreiss-Oliger dissipation with an amplitude of $0.9$ to the time-integration equations so as to damp out spurious high-frequency unstable modes of the solution. The non-dynamical variables $Q$, $\alpha$ and $\beta^r$ are integrated through the $r$-domain at every time step using the trapezoidal method. The integration for $\alpha$ is done from $r=r_{\text{max}}$, whereas that for $Q$ and $\beta^r$ are done from $r=0$. Specifically, for smoothness we assume $Q$ to be an even function of $r$ and  vanish at $r=0$, and $\beta^r$ an odd function of $r$. The boundary conditions for both dynamical and non-dynamical variables are imposed at every time step. We shall choose three sets of $c_{i}$'s, GEJ1, GEJ2, and NC, where NC denotes the choice, $c_{13} = 0, \; c_2 = 2 c_{14} = 2.0 \times 10^{-7}$,  which satisfies the constraints of Eq.(\ref{2.2}). For all three sets, the aether field is stable throughout and beyond the collapse of the scalar field to the central region. During the collapsing process, our code converges in a 2nd-order manner in line with the designed order of convergence of the numerical schemes. We further validate our code by reproducing the results of  \cite{GEJ07} for the parameter sets of GEJ1 and GEJ2. Different boundary conditions for $Q$ and $\beta^r$ at $r=0$ or $r=\Delta r$ are tested, and we find  that different boundary treatments do not affect the behavior of the PDE system in the bulk of the $r$-domain.  In our simulations for all the cases, the scalar field splits into two pieces, with one collapsing under its self-gravity toward $r=0$ and the other traveling to $r\to\infty$ (Fig.~\ref{fig:scalar}). As the collapsing piece reaches the central region, we see the formation of the apparent, spin-0 and dynamical universal horizons at finite areal radii.

\begin{figure*}
\minipage{0.33\textwidth}
  \includegraphics[width=\linewidth]{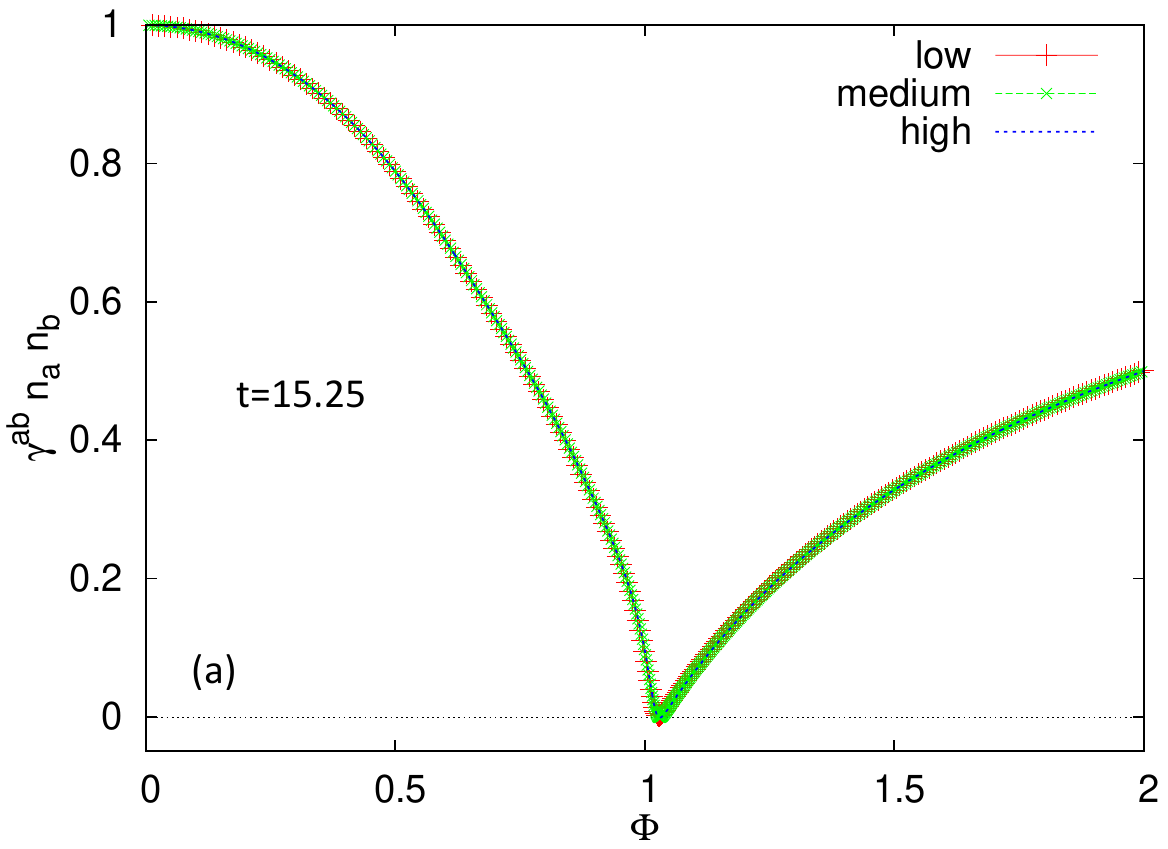}
\endminipage\hfill
\minipage{0.33\textwidth}
  \includegraphics[width=\linewidth]{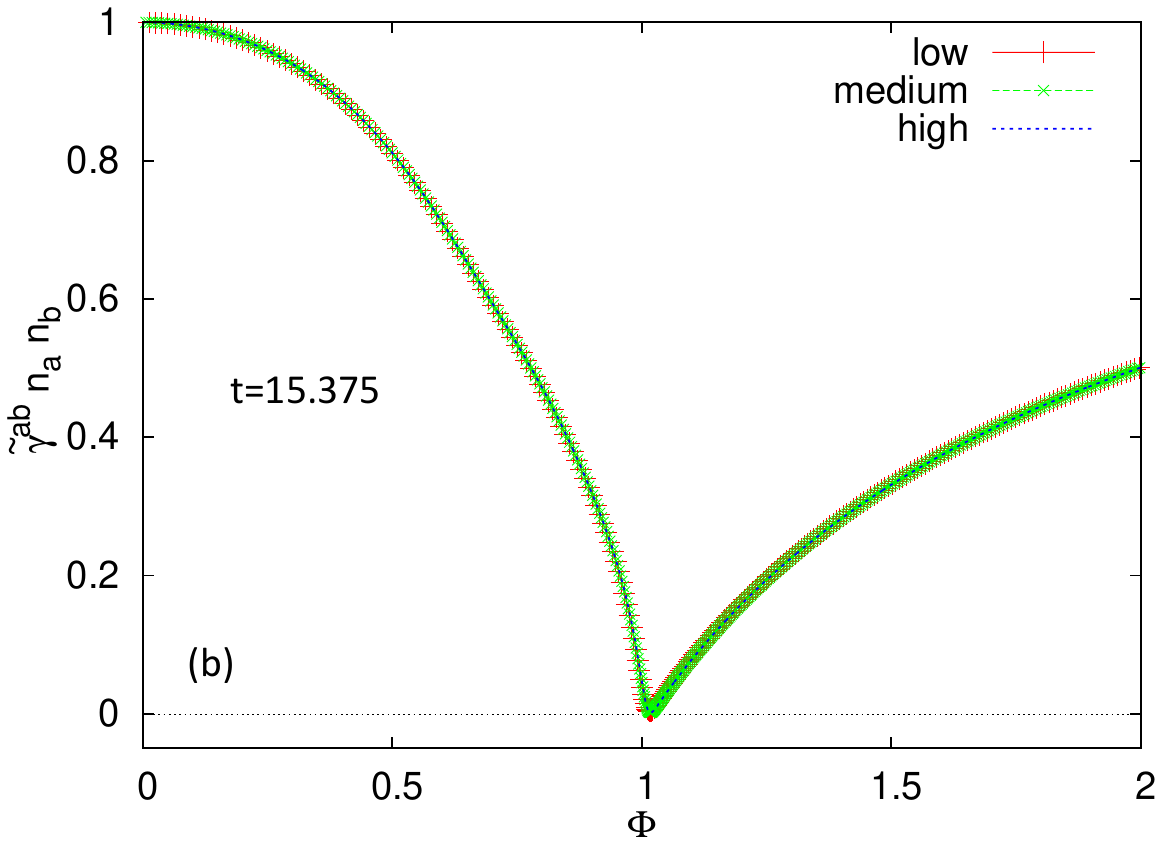}
\endminipage\hfill
\minipage{0.33\textwidth}%
  \includegraphics[width=\linewidth]{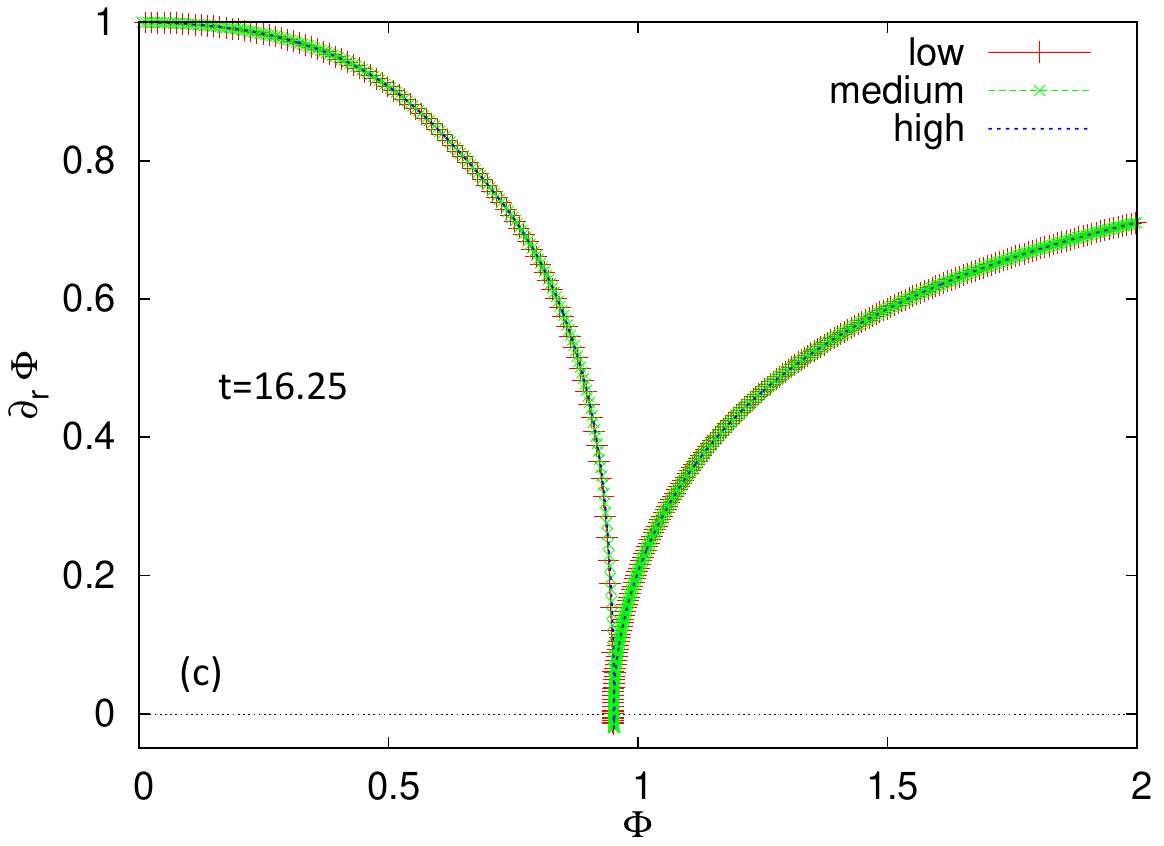}
\endminipage
\caption{Formation of (a) AH, (b) S0H, and (c) dUH for  GEJ1  at the respective times indicated in each panel. The almost complete overlap of the curves obtained from simulations with low, medium and high resolutions show that the system has almost completely converged at the low resolution of this study. \label{fig:GEJ1horizons}}
\end{figure*}

\begin{figure}
\includegraphics[width=\linewidth]{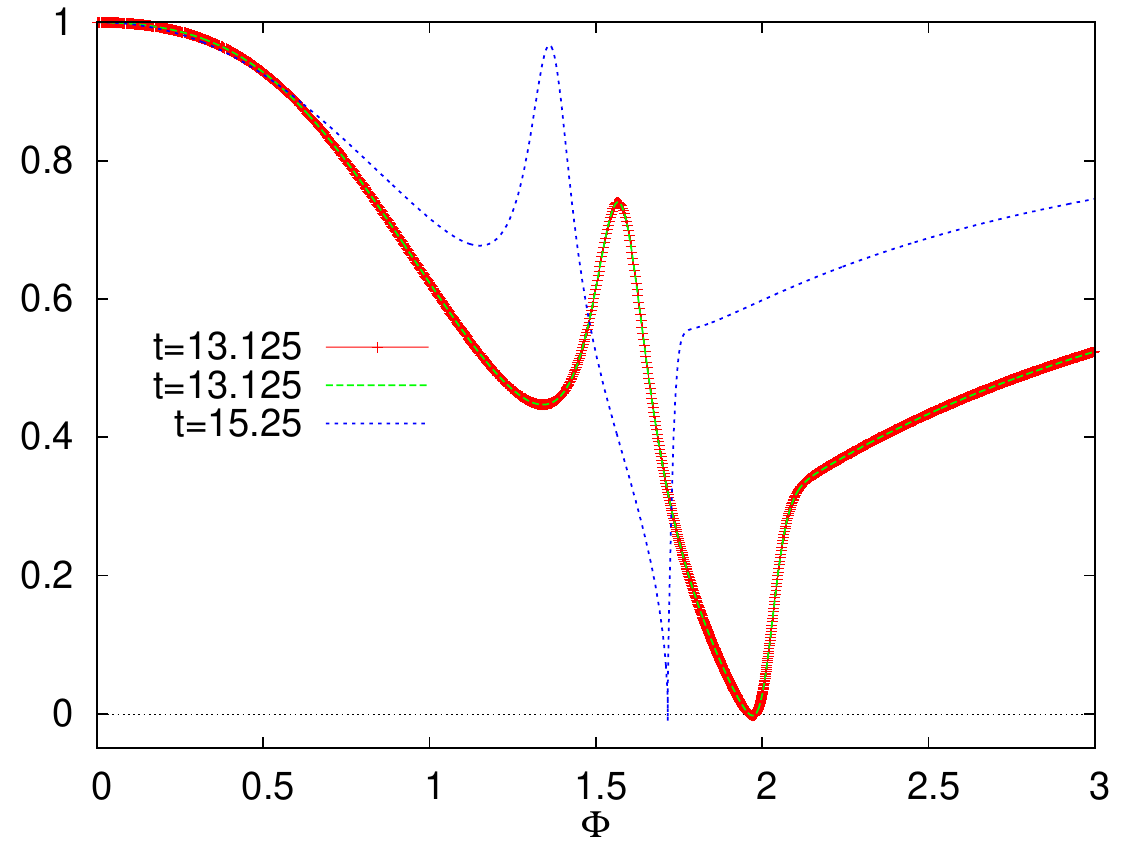}%scale=0.59]{fig2.pdf} 
\caption{Formation of  AH, S0H and dUH   for  GEJ2  at the respective times indicated in the legend. The red line with crosses represents the profile for $\gamma^{ab} n_a n_b$, the dashed green line for $\tilde\gamma^{ab} n_a n_b$, and the dashed blue line for $\partial_r\Phi$.  
 \label{fig:GEJ2horizons}}
\end{figure}

\begin{figure*}
\minipage{0.33\textwidth}
  \includegraphics[width=\linewidth]{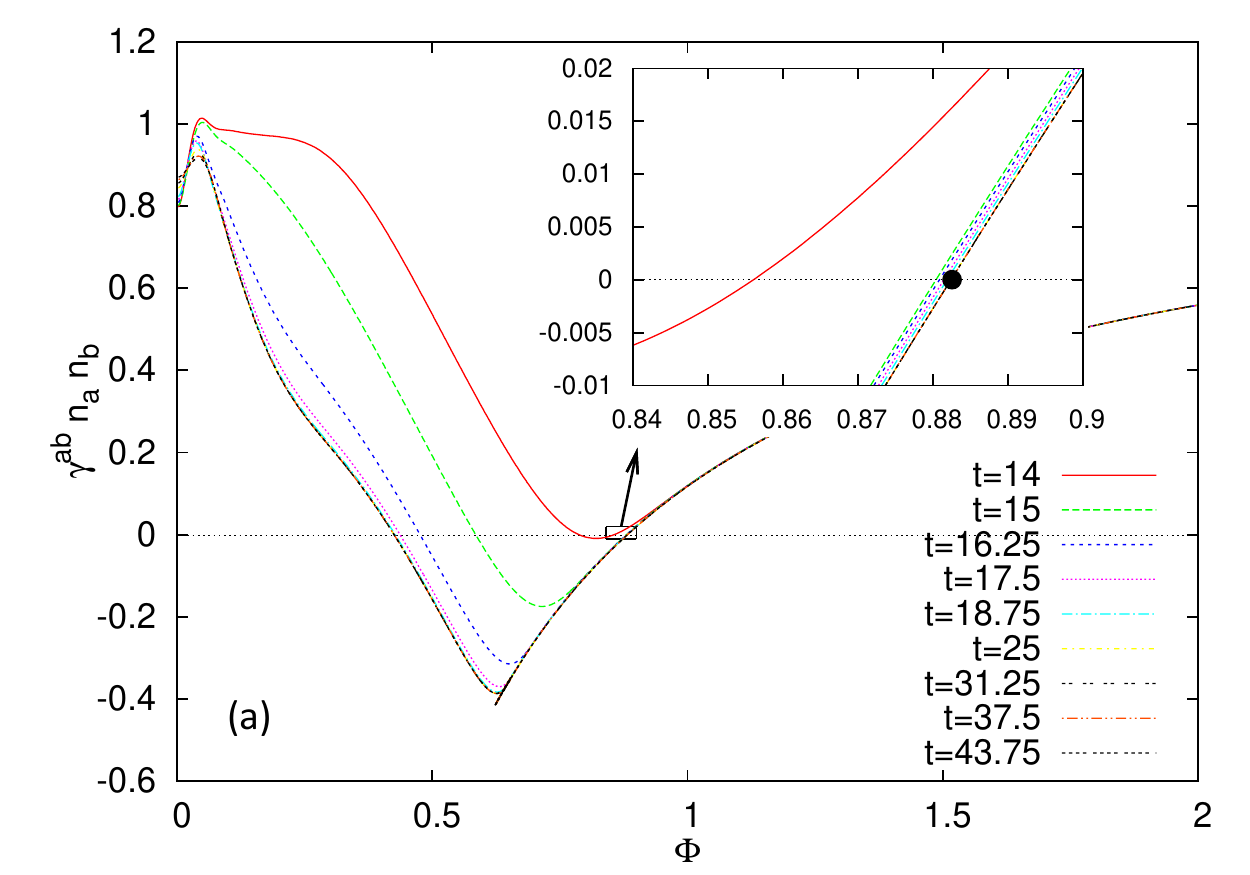}
\endminipage\hfill
\minipage{0.33\textwidth}
  \includegraphics[width=\linewidth]{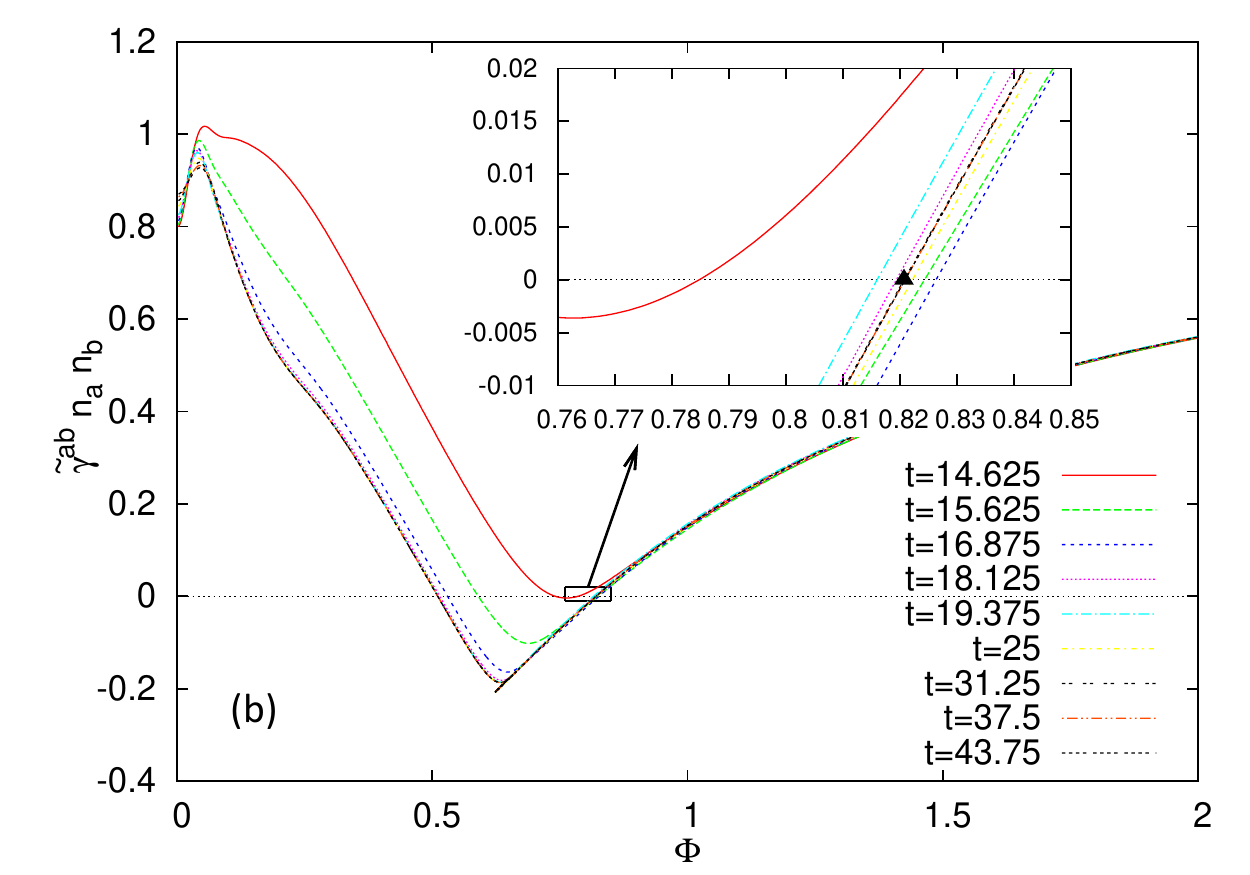}
\endminipage\hfill
\minipage{0.33\textwidth}%
  \includegraphics[width=\linewidth]{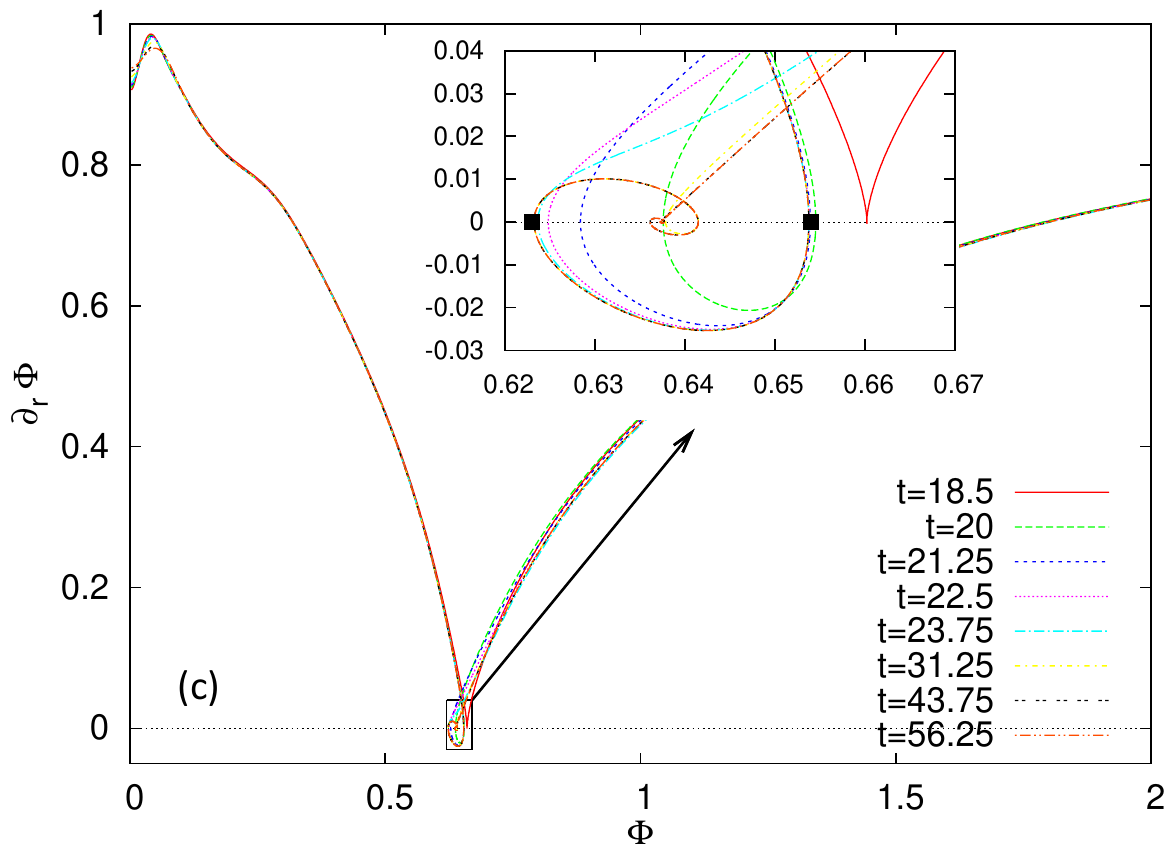}
\endminipage
\caption{Locations of (a) AH (black dot in inset), (b) S0H (black triangle in inset), and (c) dUHs (black squares in inset)  for  NC. The red line in each plot indicates the profiles shortly after the respective horizons form.}\label{fig:NChorizons}
\end{figure*}

%In our simulations for all three cases, the scalar field wave packet splits into two pieces, with one collapsing toward $r=0$ and the other traveling to $r\to\infty$.   As the collapsing piece reaches the central region, we see the formations of the AH, S0H and dUH at finite areal radii. 

Fig.~\ref{fig:GEJ1horizons} shows the profiles of $\gamma^{ab}n_a n_b$, $\tilde\gamma^{ab} n_a n_b$ and $u_a k^a (=\partial_r\Phi)$ of  GEJ1 shortly after the respective horizons  are  formed. The finite areal radii of these horizons are robust with respect to the resolutions used in this study, indicating that the system has almost completely converged at $\Delta r=0.0125$, i.e., the low resolution (Fig.~\ref{fig:GEJ1horizons}). From tests carried out using $r_{\text{max}}=80,320$ at the medium resolution, we also see that the results are robust with respect to the size of the $r$-domain. At $t=16.25$, a dUH forms at $r\approx 1.40 \; (\Phi\approx 0.95)$.  

For GEJ2, we track the collapsing process using our high-resolution simulation and similarly find the formation of all three horizons (Fig.~\ref{fig:GEJ2horizons}). As noted in   \cite{GEJ07}, the AH and S0H in this case coincide since $c_S^2=1$ and thus $\tilde{\gamma}^{ab}=\gamma^{ab}$. Hereafter, all results are obtained using high-resolution simulations, except for those with $r_{max}=320$.

For  NC, the AH forms at $t\approx 14$ and becomes quasi-stationary beginning at $t\approx 25$ with $\Phi\approx 0.8818$ (Fig.~\ref{fig:NChorizons}a). The S0H forms at $t\approx 14.625$ and achieves quasi-stationarity from $t\approx 31.25$ with $\Phi\approx 0.8210$ (Fig.~\ref{fig:NChorizons}b).  At $t\approx 18.5$, a dUH forms as a double root of $\partial_r\Phi$ at  $\Phi \approx 0.660\; (r \approx 2.0)$ (Fig.~\ref{fig:NChorizons}c). After that, the double root splits into two single roots, i.e. the inner (smaller $r$, larger $\Phi$) and outer (larger $r$, smaller $\Phi$) dUHs, and then the areal radius of the outer dUH decreases until it becomes almost constant at $t\approx 31.25$ with $\Phi\approx 0.6232$. The areal radius of the inner dUH becomes almost constant already at $t\approx 21.25$ with $\Phi\approx 0.6538$. At $t\approx 28.69$, an additional pair of dUHs forms outside the already existing pair and thus one of the new pair of dUHs becomes the outermost dUH. The areal radii of the new pair are between those of the old pair. At $t\approx 40.25$,  one more pair of dUHs forms outside the two pairs and thus one of the newest pair becomes the outermost dUH. The areal radii of the newest pair are between those of the second pair (Fig.~\ref{fig:NChorizons}c). As time increases, the number of such pairs of dUHs keeps increasing, and one of the newest pair becomes the outermost dUH. This demonstrates that even after the first pair of dUHs (denoted by the two black squares in Fig. 3(c)) has become stationary, the region {outside i.e. with larger $r$ (but with $\Phi$'s between the first pair of dUHs)} is still highly dynamical. It is interesting to note that  static black holes  (in the decoupling limit) also have  infinite layers of UHs \cite{BS11}. 

\begin{figure}
\includegraphics[width=\linewidth]{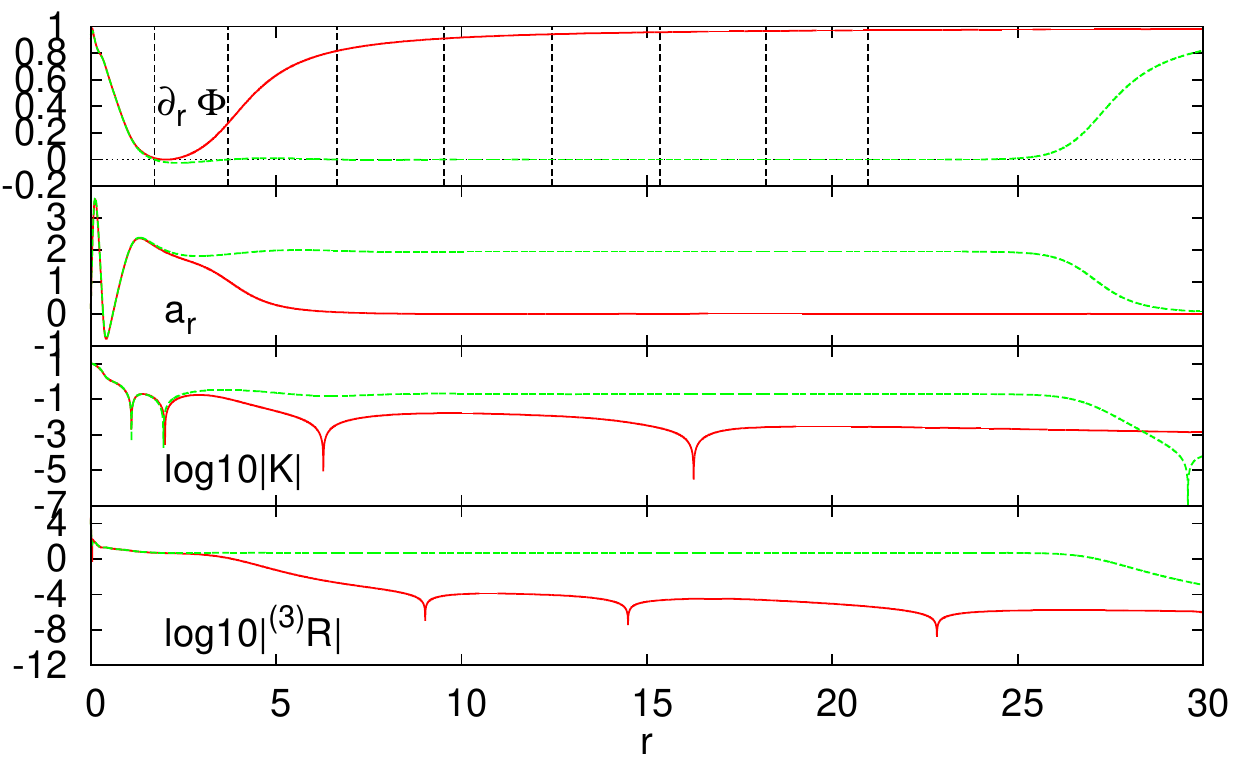} %width=0.45\textwidth,keepaspectratio]{fig5.pdf} 
\caption{ Some physical quantities vs $r$ at $t=18.5$ (solid red line) and $t=56.25$ (dashed green line) for NC. The dashed black vertical lines in the top-most panel indicate the locations of various dUHs  at $t=56.25$. 
\label{fig:NCcurv}}
\end{figure}

In Fig.~\ref{fig:NCcurv}, we show some physical quantities nearby the locations of the dUHs. While their magnitudes are much higher than those in the surrounding regions, they do not exhibit any blow-up in time, indicating that the spacetime is regular at the locations of these horizons. We note that since we have imposed the smoothness condition at $r=0$, our simulations do not show any blow-up of the curvature at $r=0$. 

\begin{figure}
\includegraphics[width=\linewidth]{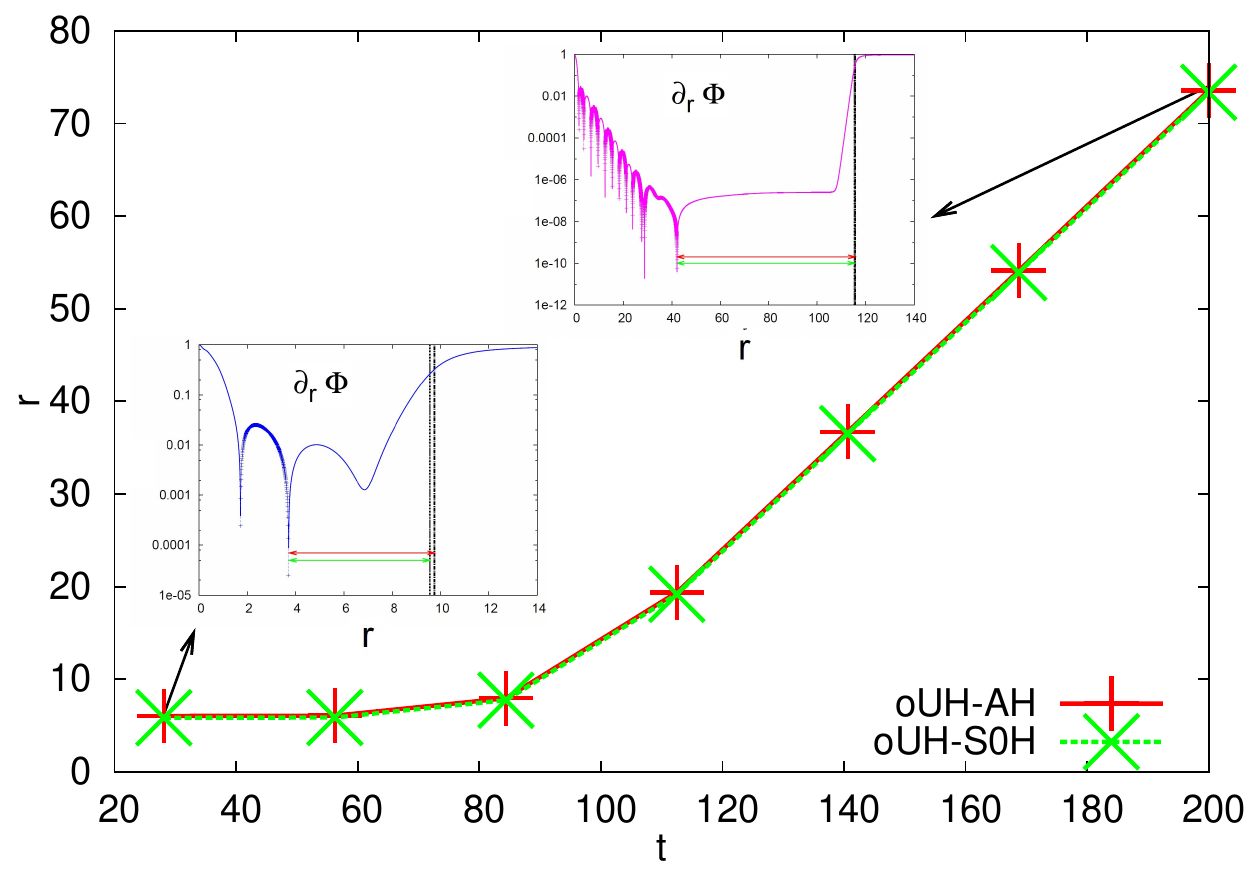} %width=0.45\textwidth,keepaspectratio]{fig6.pdf}   
\caption{Proper distance $r$ of the \textit{outermost} dUH from  AH labeled by oUH-AH and  that from S0H labeled by oUH-S0H for NC. 
\label{fig:NCcausal}}
\end{figure}

Using the result of the medium-resolution simulation with $r_\text{max}=320$, we plot the change in the proper distance of the \textit{outermost} dUH from both  AH and S0H in Fig.~\ref{fig:NCcausal}. The fact that these distances become longer and longer as time progresses indicates that the \textit{outermost} dUH is evolving into the causal boundary, even for excitations with large speeds of propagation.

\section{Conclusions}

 In GR, EHs can be formed from gravitational collapse of realistic matter, so it strongly suggests that black holes with EHs as their boundaries exist in our Universe. However, in gravitational theories with breaking Lorentz symmetry,   particles with speeds larger than that of light exist, so those EHs are  no longer the one-way membranes to such particles, as they can cross those boundaries and escape  to infinity, even initially they are trapped inside them.  Instead, now the black hole boundaries are defined by UHs. Therefore, astrophysically it is important to show UHs can be also formed from gravitational collapse of realistic matter, so even with respect to these particles black holes  also exist in our Universe    \cite{SAM14,TWSW15,BCCS16}. 

In this paper, we have numerically studied the gravitational collapse of a massless scalar field with spherical symmetry in $\ae$-theory, and shown explicitly that all three kinds of horizons, {\em apparent, spin-0 and dynamical universal},  can be formed from gravitational collapse, by considering three representative sets, GEJ1, GEJ2, and NC, of the free parameters $c_i$'s. In the cases of GEJ1 and GEJ2,  the collapse finally settles down to the regular static black holes found numerically in  \cite{EJ06}, although none of these two cases satisfies the constraints of Eq.(\ref{2.2}). Also in the  case of NC, which satisfies Eq.(\ref{2.2}), all three kinds of horizons are formed, and the spacetime in the neighborhoods of  these horizons is well-behaved and  regular, while the spacetime outside the apparent and spin-0 horizons soon settles down to a static configuration.

\section*{Acknowlodgements}

We would like to thank D. Garfinkle and S. Sibiryakov for valuable suggestions and comments. S.M. thanks Baylor University for hospitality. 
This work is supported in part by   the National Natural Science Foundation of China (NNSFC), Grant Nos. 11375153 and 11675145. 
The work of S.M. is supported by JSPS KAKENHI Grant Nos. JP17H02890, JP17H06359, and by WPI, MEXT, Japan.

\end{document}